\documentclass[12pt]{revtex4}
\usepackage{dcolumn}
\usepackage{graphicx}
\usepackage{amsmath}
\usepackage{amsfonts}
\usepackage{amssymb}
\usepackage{psfrag}
\usepackage{wrapfig}
\usepackage{subfigure}
\usepackage{makeidx}
\usepackage{bm}
\usepackage{epsf}
\usepackage{hyperref}

\begin{document}

\title{Spontaneous Symmetry Breaking Vacuum Energy in Cosmology}

\author{ Kang Zhou$^{1}$, Rui-Hong Yue$^{2}$\footnote{Email:yueruihong@nbu.edu.cn},
Zhan-Ying Yang$^{1}$\footnote{Email:zyyang@nwu.edu.cn} and De-Cheng Zou$^{1}$}
\affiliation{ $^{1}$Department of Physics, Northwest University, Xi'an, 710069, China\\
$^{2}$Faculty of Science, Ningbo University, Ningbo 315211, China}

\date{\today}

\begin{abstract}
\indent

In present paper the gravitational effect of spontaneous symmetry breaking vacuum
energy density is investigated  by subtracting the flat space-time contribution from the energy in the curved space-time.
We found that the remain effective energy-momentum tensor is too small
to cause the acceleration of the universe although it satisfies the characteristic of the dark
energy. However it could provide a promising
explanation to the puzzle why the gravitational effect produced by the huge symmetry
breaking vacuum energy in the electroweak theory has not been observed, since it has a
sufficient small value (smaller than the observed cosmic energy density by a factor $10^{32}$).
\end{abstract}



\maketitle

\section{Introduction\label{intro}}
\indent

The observational data of Supernovae Type Ia (SN Ia) accumulated by the year 1998 have shown that
the present universe is accelerating \cite{Riess:1998}, \cite{Perlmutter:1999}. The source
for this late-time cosmic acceleration was dubbed "dark energy". Despite many years of
research \cite{Peebles:2003}, \cite{Padmanabhan:2003}, \cite{Copeland:2006}, \cite{Silvestri:2009},
\cite{Caldwell:2009}, its origin has not been identified yet. Dark energy is distinguished from
ordinary matter species in the sense that it has a negative pressure \cite{Peebles:2003}.
The simplest candidate for dark energy is the so-called cosmological constant $\Lambda$,
whose energy density remains constant. From the viewpoint of particle physics, the cosmological
constant appears as vacuum energy density. The  cosmological constant problem had been
investigated long before the discovery of dark energy in 1998. In 1989, Weinberg pointed out
that possible candidates for vacuum energy density can be the summing of zero-point energies
of all normal modes of fields up to a cut-off scale of the momentum, or the vacuum expectation
value of the energy density in a certain state which has spontaneous symmetry breaking
\cite{Weinberg:1989}. If we believe general relativity up to the Planck scale and take
$\Lambda\simeq(8\pi G)^{-1/2}$ as the cut-off scale, the zero-point energy would give
$\rho_{vac}\simeq 2\times10^{71}GeV^4$. On the other hand,
the vacuum expectation value of the energy density in  electroweak theory would give $\rho_{vac}\simeq-g(300GeV)^4$
while $g$ is the coefficient of the $(\phi^\dag\phi)^2$ term in Higgs Lagrangian. Even for $g$
as small as $\alpha^2$ with $\alpha$ the fine structure constant, it would yield $|\rho_{vac}|\simeq10^6 GeV^4$.
Hence, two candidates are all much larger than the observed value of the total energy density of the
present universe: $\rho\simeq 10^{-47} GeV^4$ \cite{Weinberg:1989}. It is problematic.
Furthermore, a nature puzzle would be  why the gravitational effect produced by such huge vacuum
 energy density has not been detected.

Recently, Maggiore proposed a new method to calculate the  zero-point fluctuations
of quantum fields in cosmology \cite{Maggiore:2011}. In the Hamiltonian formulation of the
general relativity, the energy associated to an asymptotically flat space-time with
metric $g_{\mu\nu}$ is related to the Hamiltonian $H_{GR}$ by $E=H_{GR}[g_{\mu\nu}]
-H_{GR}[\eta_{\mu\nu}]$, where the subtraction of the flat-space contribution is
necessary to get  rid of an otherwise divergent boundary term \cite{Arnowitt:1962}.
The classical result indicates that the energy associated to flat space-time does not
gravitate. The definition of standard ADM mass is based on such
principle. Maggiore applied this principle to study the effective zero-point energy,
proposing that their contribution to the dynamic of the universe is
 obtained by computing the vacuum energy of quantum fields in a FRW space-time and
 subtracting from it the flat space-time contribution. Although the remained energy
 density after doing cut-off procedure is not compatible with the present
 observations of dark energy, his concept provided a new
credible method to calculate the vacuum energy density.

We apply such idea to study the spontaneous symmetry breaking vacuum energy
density in an electroweak theory in the background of a spatially flat FRW space-time, and explore
 its gravitational effect. After subtracting the energy density in the flat space-time,
 the effective energy-momentum tensor satisfies the characteristic of the dark
 energy. However, we find that the value of such vacuum energy density is much smaller than
 the observed cosmic energy density by a factor $10^{32}$, therefore it could not be regarded
 as a candidate for dark energy. Nevertheless, this effective vacuum energy density with so
 small value provides a promising interpretation why the gravitational
 effect of the symmetry breaking vacuum energy have not been detected, since its effect
 on gravity can be neglect while comparing  to the observed cosmic energy.

This paper is organized as follows. The subject of Sec \ref{22s} is to calculate the
effective symmetry breaking vacuum energy density by Maggiore's method.  In Sec \ref{33s},
 we investigate the gravitational effect produced by such vacuum energy density, and try
  to explain why the symmetry breaking vacuum energy has not been observed. Sec \ref{44s}
   is devoted to discussions and conclusions.

\section{spontaneous symmetry breaking vacuum energy density\label{22s}}
\indent

The WMAP 5-year data constrain that the contribution of the curvature of  present universe
to be only $-0.0175\sim0.0085$ \cite{Komatsu:2009}, while the total contribution which
contains matter, dark energy and curvature is equal to $1$. Hence, for simplicity, we
consider only the case of a spatially flat FRW space-time. In the preferred coordinate
system, the metric takes the
form \cite{Robertson:1935}, \cite{Walker:1936}
\begin{eqnarray}
ds^2=dt^2-a^2(t)\delta_{ik}dx^idx^k.\label{eq:01a}
\end{eqnarray}
It is convenient to introduce the conformal time $\eta\equiv\int\frac{dt}{a(t)}$ instead of the cosmic time $t$. With this new coordinate  the metric.~(\ref{eq:01a}) becomes into
\begin{eqnarray}
ds^2=a^2(\eta)\eta_{\mu\nu}dx^\mu dx^\nu,\label{eq:02a}
\end{eqnarray}
and it is obvious that the metric is conformally equivalent to the Minkowski metric $\eta_{\mu\nu}$.
Let us consider the effective vacuum energy density with spontaneous symmetry breaking in such a space-time.
 Since we only concentrate  the present universe, the electroweak phase transition in the early universe will be neglected.

In the electroweak theory, the Lagrangian density of Higgs scalar fields
takes the form of \cite{Higgs:1964}, \cite{Weinberg:1967}
\begin{eqnarray}
{\cal L}=\frac{1}{2}g^{\mu\nu}\phi^\dag_{,\mu}\phi_{,\nu}-\frac{1}{2}\mu^2\phi^\dag
\phi-\frac{\lambda}{4}(\phi^\dag \phi)^2\label{eq:03a}
\end{eqnarray}
with $\mu^2<0$, $\lambda>0$. Putting the Higgs field into the above conformal flat space-time,
the action reads
\begin{eqnarray}
S&=&\int d^3x d\eta [\frac{1}{2}a^2{\phi^\dag}'\phi'-\frac{1}{2}a^2\nabla\phi^\dag\nabla\phi\nonumber\\
&-&\frac{1}{2}\mu^2 a^4\phi^\dag\phi-\frac{\lambda}{4}a^4(\phi^\dag \phi)^2+ba^4\xi R\phi^{\dag}\phi],\label{eq:04a}
\end{eqnarray}
where the prime denotes derivatives with respect to the conformal time $\eta$.
The condition that $\phi_{;\mu}=\phi_{,\mu}$ for scalar fields has been used.
The term $b\xi R\phi^{\dag}\phi$ describes the interaction between the Higgs field and the
gravitational field \cite{Buchbinder:1986}. Here $b$ is a positive number, $\xi$ is
the parameter of coupling of Higgs and gravitational fields, and $R$ is the scalar
curvature of the external gravitational field.
In the following calculation, the term $ba^4\xi R\phi^{\dag}\phi$ will be absorbed into the mass via ${\hat{\mu}}^2=\mu^2-2b\xi R$,
since in the present homogeneous, isotropic universe the  $R=6(\ddot{a}/a+{\dot{a}}^2/a^2)$ keeps  small and almost invariant. Meanwhile  $b\xi$ is a small coupling coefficient and ensures  ${\hat{\mu}}^2$ is still negative.

Introducing auxiliary fields as $\varphi=a(\eta)\phi$ and $\varphi^\dag=a(\eta)\phi^\dag$,
we can rewrite the action Eq.~(\ref{eq:04a}) in terms of $\varphi$ and $\varphi^\dag$ as
\begin{eqnarray}
S&=&\int d^3x d\eta [\frac{1}{2}{\varphi^\dag}'\varphi'-\frac{1}{2}\nabla\varphi^\dag\nabla\varphi\nonumber\\
&-&\frac{1}{2}({\hat{\mu}}^2a^2-\frac{a''}{a})\varphi^\dag\varphi-\frac{\lambda}{4}(\varphi^\dag\varphi)^2].\label{eq:05a}
\end{eqnarray}
That is, the action of fields $\phi$ and $\phi^\dag$ in curved FRW space-time equivalently to
the ones of  $\varphi$ and $\varphi^\dag$ in flat Minkowski space-time.
Thus, the  Lagrangian density of new system can be written  as
\begin{eqnarray}
{\cal\widetilde{L}}&=&\frac{1}{2}{\varphi^\dag}'\varphi'-\frac{1}{2}\nabla\varphi^\dag\nabla\varphi 
-\frac{1}{2}({\hat{\mu}}^2a^2-\frac{a''}{a})\varphi^\dag\varphi-\frac{\lambda}{4}(\varphi^\dag\varphi)^2,\label{eq:06a}
\end{eqnarray}
and the Hamiltonian density is
\begin{eqnarray}
{\cal \widetilde{H}}&=&\frac{\partial {\cal\widetilde{L}}}{\partial\varphi'}\varphi+\varphi^\dag\frac{\partial {\cal\widetilde{L}}}{\partial{\varphi^\dag}'}-{\cal \widetilde{L}}\nonumber\\
&=&\frac{1}{2}{\varphi^\dag}'\varphi'+\frac{1}{2}\nabla\varphi^\dag\nabla\varphi\nonumber\\
&+&\frac{1}{2}({\hat{\mu}}^2a^2-\frac{a''}{a})\varphi^\dag\varphi+\frac{\lambda}{4}(\varphi^\dag\varphi)^2.\label{eq:07a}
\end{eqnarray}

The vacuum was defined as an eigenstate of the Hamiltonian with the lowest possible energy. In the case under consideration the Hamiltonian depends explicitly on time and thus does not possess time-independent eigenvectors. Nevertheless, given a particular moment of time $\eta_0$ we can still define the instantaneous vacuum $|0_{\eta_0}\rangle$ as the lowest-energy state of the Hamiltonian $H(\eta_0)$. The classical Hamiltonian  has an extrema at the moment $\eta_0$ if $\varphi'\Big|_{\eta_0}={\varphi^\dag}'\Big|_{\eta_0}=0$, $\nabla\varphi\Big|_{\eta_0}=\nabla\varphi^\dag\Big|_{\eta_0}=0$ and
$\frac{\partial V}{\partial\varphi}\Big|_{\eta_0}=\frac{\partial V}{\partial\varphi^\dag}\Big|_{\eta_0}=0$,
with $V(\varphi^\dag,\varphi)=\frac{1}{2}({\hat{\mu}}^2a^2-\frac{a''}{a})\varphi^\dag\varphi
+\frac{\lambda}{4}(\varphi^\dag\varphi)^2$. For ${\hat{\mu}}^2<0$, $\lambda>0$, the solution of Higgs fields  satisfies
\begin{eqnarray}
\varphi^\dag(\eta_0)\varphi(\eta_0)=\frac{-{\hat{\mu}}^2a(\eta_0)^2+a''(\eta_0)/a(\eta_0)}{\lambda}.\label{eq:08a}
\end{eqnarray}
Consequently, the vacuum expectation value of the energy density in classical approximation is given by
\begin{eqnarray}
{\cal\widetilde{H}}_{min}(\eta_0)=-\frac{(-{\hat{\mu}}^2a(\eta_0)^2+a''(\eta_0)/a(\eta_0))^2}{4\lambda}.\label{eq:09a}
\end{eqnarray}
Thus  the physical vacuum  energy regarded as the total  minimum energy of the system  is
\begin{eqnarray}
E_{vac}(\eta_0)&=&\int d^3x  {\cal\widetilde{H}}_{min}(\eta_0)\nonumber\\
&=&\int d^3x  \sqrt{-g(\eta_0)}[-\frac{(-{\hat{\mu}}^2+\frac{a''(\eta_0)}{a(\eta_0)^3})^2}{4\lambda}],\label{eq:10a}
\end{eqnarray}
which leads to a vacuum energy density
\begin{eqnarray}
\rho_{vac}(\eta_0)=-\frac{(-{\hat{\mu}}^2+a''(\eta_0)/a(\eta_0)^3)^2}{4\lambda}.
\end{eqnarray}
Obviously, this conclusion is appropriate for arbitrary moment $\eta$.
Since $d^3x d\eta \sqrt{-g}$ is an invariant volume element, it is equivalent to
the element $d^4x\sqrt{-g}=d^4x$ in the Minkowski space-time. That is,
we only need to compare the vacuum energy density $\rho_{vac}$ and $\rho'_{vac}$ while
comparing  the total vacuum energy $E_{vac}$ in the FRW space-time and $E'_{vac}$ in
the Minkowski space-time.

Follow Maggiore's viewpoint, the vacuum energy in the flat space-time has no
contribution to the dynamic of gravitational system,  which should be subtracted
from the vacuum energy density in the FRW space-time.
The vacuum energy density in the electroweak theory in the Minkowski space-time
is $\rho'_{vac}=-\frac{\mu^4}{4\lambda}$. Hence, we get the effective vacuum energy density in cosmology is
\begin{eqnarray}
\rho_{eff}=\rho_{vac}-\rho'_{vac}=\frac{2\mu^2(2b\xi R+a''/a^3)-(2b\xi R+a''/a^3)^2}{4\lambda}.\label{eq:11a}
\end{eqnarray}
It illustrates that  the effective vacuum energy density $\rho_{eff}$ depends
on the value of parameters $\mu^2$ and $\lambda$. Since the Higgs particle has not been found yet,
these two values remain unclear. However, the value of $\sqrt{-\mu^2/\lambda}$ which is the
vacuum expectation value of the Higgs field in the electroweak theory has been obtained
as $\langle\phi\rangle\simeq300GeV$ \cite{Weinberg:1989}. If we assume the value of the
the term $(a''/a^3)^2$ is very low and neglect it, the value of $\rho_{eff}$ could be calculated with given $a(t)$. The rationality of such an assumption will be discussed in the next section.

In particle physics, it is licit to adding a term $V_0$ to the Lagrangian Eq.~(\ref{eq:03a}) with $V_0$
a constant, the physical effect in scattering processes would not be modified. Clearly,
the contribution of $V_0$ has been canceled in present approach.
Hence, the result about the effective vacuum energy density is independent on  $V_0$.

The energy-momentum tensor of scalar fields in curved space-time is \cite{Parker:1974}, \cite{Fulling:1974}
\begin{eqnarray}
T_{\mu\nu}=\phi^\dag_{,\mu}\phi_{,\nu}-g_{\mu\nu}(\frac{1}{2}g^{\rho\sigma}\phi^\dag_{,\rho}\phi_{,\sigma}-V(\phi^\dag,\phi)).\label{eq:12a}
\end{eqnarray}
For the vacuum state, which implies $\varphi'\Big|_{\eta_0}={\varphi^\dag}'\Big|_{\eta_0}=0$, $\nabla\varphi\Big|_{\eta_0}=\nabla\varphi^\dag\Big|_{\eta_0}=0$
and $\varphi^\dag\varphi\Big|_{\eta_0}=\frac{-{\hat{\mu}}^2a^2+a''/a}{\lambda}$, it is straightforward to obtain
the non-vanishing components  $T_i^i=T_0^0=\rho_{vac}$ and ${T'}_i^i={T'}_0^0=\rho'_{vac}$.
Subtracting the flat space-time contribution from the energy-momentum tensor in curved space-time,
 we get $\rho_{eff}=-p_{eff}$. Consequently, the energy-momentum tensor of the vacuum  with  symmetry breaking satisfies the characteristic of the dark energy. However, the expression of such energy density Eq.~(\ref{eq:11a}) is dependent on time, which could not be regarded as the cosmological constant.

\section{gravitational effect\label{33s}}
\indent
In this section, we apply such vacuum energy-momentum tensor to the expanding universe,
and consider its physical effect to the gravitational system. The Einstein equation
$G_{\mu\nu}\equiv R_{\mu\nu}-\frac{1}{2}g_{\mu\nu}R=8\pi GT_{\mu\nu}$ in the spatially flat FRW space-time gives
\begin{eqnarray}
&H^2&=\frac{8\pi G}{3}\rho,\label{eq:13a}\\
&\dot{\rho}&+3H(\rho+p)=0,\label{eq:14a}
\end{eqnarray}
where $H\equiv\dot{a}/a$ is the Hubble parameter. The dot represents a derivative with respect to the cosmic time $t$. We ignore the contribution of the radiation, and regard the present universe as an ideal fluid which contains the non-relativistic matter (include the dark matter) satisfying  $p_M=0$ and the dark energy satisfying $p_{DE}=\omega_{DE}\rho_{DE}$. From the combined analysis of SN Ia, CMB, and BAO, the WMAP group obtained the bound $-1.097<\omega_{DE}<-0.858$ \cite{Komatsu:2009}. Hence, We assume $\omega_{DE}=-1$, then Eq.~(\ref{eq:13a}) and Eq.~(\ref{eq:14a}) take the form of
\begin{eqnarray}
&H^2&=\frac{8\pi G}{3}(\rho_M+\rho_{DE}),\label{eq:15a}\\
&\dot{\rho}_M&+\dot{\rho}_{DE}+3H\rho_M=0.\label{eq:16a}
\end{eqnarray}
Remark that the matter and the dark energy  do not  satisfy the energy-momentum conservation
equation Eq.~(\ref{eq:14a}) respectively. Instead, we only require the summation of them to
satisfy the total conservation equation Eq.~(\ref{eq:16a}). In other words, we permit the transformation between the matter and the dark energy since particles could be created from the vacuum in quantum field theory in curved space-time \cite{Hawking:1975}, \cite{Unruh:1976}. From Eq.~(\ref{eq:15a}) and Eq.~(\ref{eq:16a}), we get
\begin{eqnarray}
\dot{H}=4\pi G\rho_{DE}-\frac{3}{2}H^2=\frac{3}{2}H^2(\Omega_{DE}-1),\label{eq:17a}
\end{eqnarray}
where $\Omega_{DE}\equiv\rho_{DE}/(\rho_{DE}+\rho_M)$. If the expression of $\rho_{DE}$ is given,
one can obtain the exact solution of $H(t)$ by solving Eq.~(\ref{eq:17a}). On the other hand, we can calculate the present value of $\dot{H}_0$. The present Hubble parameter is recognized as $H_0=2.1332h\times10^{-42}GeV$ with $h=0.72\pm0.08$ \cite{Freedman:2001}, and the combined data analysis have provided the
following constraint for the present density parameter of dark energy which is $\Omega_{DE(0)}=0.726\pm0.015$ \cite{Komatsu:2009}. Thus, we have $\dot{H_0}=\frac{3}{2}H_0^2(\Omega_{DE(0)}-1)=-9.6955\times10^{-85}GeV^2$.

We now consider whether $\rho_{DE}$ could be the symmetry breaking vacuum energy density $\rho_{eff}$. From Eq.~(\ref{eq:11a}), we know the energy density $\rho_{eff}$ is negative. Combining Eq.~(\ref{eq:15a}) and Eq.~(\ref{eq:16a}), we get
\begin{eqnarray}
\frac{\ddot{a}}{a}=-\frac{4\pi G}{3}(\rho_{DE}+\rho_M+3p_{DE}).\label{eq:18a}
\end{eqnarray}
If the density of the dark energy $\rho_{DE}$ is equal to $\rho_{eff}$, $\ddot{a}$ is evidently negative and is to
indicate an decelerate expansion of the present universe. It is of course problematic. As we will show, the contribution of $\rho_{eff}$ to $\rho_{DE}$ is small enough to be ignored. It is worthy to point out that the term $(2b\xi R+a''/a^3)^2$ in Eq.~(\ref{eq:11a}) is assumed to be sufficiently small and to  be neglected, to  have $\rho_{eff}=\frac{2\mu^2(2b\xi R+a''/a^3)}{4\lambda}$. In fact, using the relation $d\eta=dt/a(t)$ and $R=6(\ddot{a}/a+{\dot{a}}^2/a^2)$ we get
\begin{eqnarray}
\rho_{eff}=\frac{\mu^2}{2\lambda}(12b\xi+1)(\frac{\ddot{a}}{a}+\frac{{\dot{a}}^2}{a^2}).\label{eq:19a}
\end{eqnarray}
We can neglect the term $12b\xi$ since $b\xi$ is a small coupling coefficient. Substituting $H\equiv\dot{a}/a$ into Eq.~(\ref{eq:19a}) we obtain
\begin{eqnarray}
\rho_{eff}=\frac{\mu^2}{2\lambda}(2H^2+\dot{H}).\label{eq:20a}
\end{eqnarray}
Note that $\rho_{eff}$ is proportional to $H^2(t)$ since $\dot{H}=\frac{3}{2}H^2(\Omega_{DE}-1)$.
Using values $\sqrt{-\mu^2/\lambda}\simeq300GeV$, $H_0=2.1332h\times10^{-42}GeV$
and $\dot{H_0}=-9.6955\times10^{-85}GeV^2$, we get $\rho_{eff}\simeq-1.69\times10^{-79}GeV^4$.
It is smaller than the observed cosmic energy density $\rho\simeq10^{-47}GeV^4$
by a factor $10^{32}$, therefore it could be ignored. Hence, although the symmetry
breaking vacuum energy may have contribution to the dark energy, such contribution
is too small  to explain the observed value of the dark energy.

A simple calculation shows that $a''/a^3=2H^2+\dot{H}=3.7485\times10^{-84}GeV^2$, which is very small.
It is reasonable to neglect the term $(2b\xi R+a''/a^3)^2$ while calculating $\rho_{eff}$.
Unless, the mass of the Higgs particle $m=\sqrt{-2\mu^2}$ is
$m\simeq10^{-42}GeV$. It is clearly impossible since Higgs particles with so small
 mass should be found easily.

On the other hand, despite our result could not be used to solve the dark energy problem,
the small value of $\rho_{eff}$ can be used to explain the puzzle why the symmetry breaking
vacuum energy has not been observed. If the vacuum energy density in the flat space-time has
not been subtracted, such energy density with the value $|\rho_{vac}|\simeq10^6 GeV^4$
(for the coupling constant as small as $\alpha^2$), which is larger than the observed cosmic
energy density $\rho\simeq10^{-47}GeV^4$ by a factor $10^{53}$, should produce remarkable effect
on gravity. The gravitational effect of the vacuum energy is sufficient to determine the evolution
of the universe and the contribution of the observed matter could be neglect. However, if we use
the new concept which based on the Hamiltonian formulation of the general relativity to calculate
the symmetry breaking vacuum energy, then the contribution of the vacuum energy in the flat
space-time has been subtracted, and remain a small value which is smaller than the observed
cosmic energy density by a factor $10^{32}$. Obviously, the gravitational contribution of the
symmetry breaking vacuum energy is difficult to detect.

There is another way  which does not require the subtraction of the flat space-time contribution to explain such problem. As mentioned above, one can correct the Lagrangian of Higgs fields by adding a term $V_0$. The absolute energy $-V_0$ would have gravitational effect. One can ensure the value of $V_0$ by arguing that: the term $-V_0$ cancels the term $\rho_{vac}$ (or $\rho'_{vac}$) and leads the symmetry breaking vacuum energy to be never observed. However, the principle for deciding the absolute value of $V_0$ is still absent, we know of no reason why $V_0$ should has such value rather than another. Adding a constant term $V_0$ to the electroweak action bears a strong resemblance to adding the cosmological constant term to the Einstein-Hilbert action. As in the dark energy problem, it is necessary to seek the origin of the constant term. Moreover, adding a constant to the Lagrangian of matter fields can be interpreted as changing the zero point of the potential term. Hence, if we calculate the vacuum energy without the subtraction of the flat space-time contribution, the result depends on the zero point of the potential. It means that the gravitational effect will be altered by changing the zero point of the potential, therefore violates our intuition. Oppositely, our explanation does not depend on the value of $V_0$ since it has been canceled. It does not require an unknown principle to determine $V_0$ and does not depend on the selection of the zero point of the potential.

\section{discussions and conclusions\label{44s}}
\indent

In this paper, we have investigated the spontaneous symmetry breaking vacuum energy density in the electro-weak theory in the spatially flat FRW space-time by a new method that subtracting the flat space-time contribution from the energy in the curved space-time. As discussed above, such method leads to a remained effective energy density which is independent on the selection of the zero point of the potential term in the Lagrangian of matter fields, therefore is consistent with our intuition. The effective energy density is too small to cause the acceleration of the universe. Therefore, it could not be a promising candidate for the dark energy, although it could has the contribution to the dark energy.

Our result shows that the symmetry breaking vacuum energy density value is proportional to $H^2(t)$. The zero-point energy density calculated by the same principle is also proportional to $H^2(t)$ \cite{Maggiore:2011}. Thus, the total vacuum energy density of quantum fields contains the zero-point energy and the symmetry breaking energy, which  depends on time. Although the value of the zero-point energy density can be modified by changing the cut-off scale to gain a suitable value of the total vacuum energy density, the time dependence of the total vacuum energy is not compatible with present observations of the dark energy. Thus, the vacuum energy density calculated by Maggiore's method could not solve the dark energy problem.

However, our result is useful to explain the puzzle that the gravitational effect of the symmetry breaking vacuum energy has not been observed. There are two problems leads the quantum field theory to be in conflict with the general relativity, one is why the gravitational effect of the huge zero-point energy density has not been observed, while another one is why the gravitational effect of the huge symmetry breaking vacuum energy density has not been detected. For the second problem, our result can provide a possible explanation to reconcile the electroweak theory with the general relativity.

{\bf Acknowledgment }
This work has been supported by the Natural Science Foundation of China
under grant No.10875060, 10975180 and 11047025.

\end{document}